\RequirePackage{fix-cm}
\documentclass[smallextended]{svjour3}       
\smartqed  
\usepackage{graphicx}
\graphicspath{{figures/}{./}}

\usepackage[ 
final,
breaklinks=true,
colorlinks=true,           
pdfborder={0 0 1},    
citecolor={blue},linkcolor={blue},urlcolor={red},
citebordercolor={1 0 0},linkbordercolor={0 0 1},urlbordercolor={1 0 0},
pdfpagemode=UseOutlines,
bookmarks=true,bookmarksopenlevel=4
]{hyperref}         
\usepackage{amssymb}
\usepackage{amsmath}
\usepackage{slashed}
\usepackage{bbm} 
\usepackage{xspace}
\usepackage{doi} 
\usepackage[numbers,sort&compress]{natbib}
\usepackage{xparse}
\NewDocumentCommand{\arxiv} %
{r [: u{ [} u{]]} }{[\href{http://arxiv.org/abs/#2}{arXiv:#2}~[#3]]}
\NewDocumentCommand{\arxivold} {r[]}{[\href{http://arxiv.org/abs/#1}{#1}]}
\NewDocumentCommand{\arXiv} %
{r [: u{ [} u{]]} }{[\href{http://arxiv.org/abs/#2}{arXiv:#2}~[#3]]}
\NewDocumentCommand{\arXivold} {r[]}{[\href{http://arxiv.org/abs/#1}{#1}]}
\newcommand{\3}{{\ss}}
\newcommand{\eg}{\emph{e.g.\xspace}}
\newcommand{\etal}{\emph{et al.\xspace}}
\newcommand{\etc}{\emph{etc.\xspace}}
\newcommand{\ie}{\emph{i.e.\xspace}}
\newcommand{\cf}{\emph{cf.\xspace}}

\newcommand{\ptyp}{p_\text{typ}}
\newcommand{\LambdaEFT}{\overline{\Lambda}_\text{EFT}}
\newcommand{\mpi}{\ensuremath{m_\pi}}
\newcommand{\EFTNoPion}{EFT($\slashed{\pi}$)\xspace}
\newcommand{\LambdaNoPion}{\overline{\Lambda}_{\slashed{\pi}}}
\newcommand{\ChiEFT}{$\chi$EFT\xspace}
\newcommand{\LambdaChi}{\overline{\Lambda}_{\chi}}
\newcommand{\MeV}{\ensuremath{\mathrm{MeV}}}
   
\newcommand{\vectorwithspace}[1]{\vec{#1}\mkern2mu\vphantom{#1}}
\newcommand{\kv}{\vectorwithspace{k}}
\newcommand{\qv}{\vectorwithspace{q}}
\newcommand{\dd}{\mathrm{d}}
\newcommand{\ii}{\mathrm{i}}

\newcommand{\N}{\mathrm{N}}

\newcommand{\calA}{\mathcal{A}}
\newcommand{\calO}{\mathcal{O}}

\newcommand{\obs}{\mathrm{Obs}}

\newcommand{\one}{\mathbbm{1}}

\newcommand{\wave}[3]{\ensuremath{{}^{#1}\mathrm{#2}_{#3}}}

\newcommand{\deintdim}[2]{\frac{\dd^{#1}\;\!\! #2}{(2\pi)^{#1}}\;}

\newcommand{\NXLO}[1]{N\ensuremath{{}^{#1}}LO\xspace}
\begin{document}

\title{What Can Possibly Go Wrong?}



\author{Harald. W.~Grie\3hammer}


\institute{Harald. W.~Grie\3hammer\at Institute for Nuclear Studies,
  Department of Physics, \\
  The George Washington University, Washington DC 20052, USA\\
  \email{hgrie@gwu.edu} }

\date{Received: date / Accepted: date}

\maketitle

\begin{abstract}
  A lot. 

\end{abstract}

\vspace*{2ex}
\begin{flushright}\emph{There is no principle built into the laws of Nature\\
    that says that theoretical physicists have to be happy.}~\cite{WeinbergNOVA}
\end{flushright}
\section{Motivation}
\label{sec:introduction}

In this hard, unbiased and objective look at some past and continuing blunders
in following Weinberg's suggestions to arrive at a comprehensive description
of Nuclear Physics using Effective Field Theories, some names and citations
are withheld to protect the innocent.
  
Scientific history is often told as inevitable and steady progress towards a
more perfect theory, with amusing asides about a few endearing follies of key
protagonists. This volume may provide a good excuse to replace the pressure of
result-oriented rigour in novelty-research articles by a few qualitative
remarks. None are original, and all are most likely standard lore. They were
triggered in part by presentations and discussions at the workshops
\textsc{The Tower of Effective (Field) Theories and the Emergence of Nuclear
  Phenomena (EFT and Philosophy of Science)} at CEA/SPhN Saclay in
2017~\cite{SaclayEPJA},
\textsc{Lattice Nuclei, Nuclear Physics and QCD -- Bridging the Gap} and
\textsc{New Ideas in Constraining Nuclear Forces} at the ECT* in 2015 and
2018, respectively,
and by the ``Folk
Theorem'' of Effective Field Theories (EFTs), originally formulated by
Weinberg in 1979~\cite{Weinberg:1978kz} and here quoted in the 1997
version~\cite{Weinberg:1996kw}:

\begin{quote}
\emph{When you use quantum field theory to study low-energy phenomena, then
according to the folk theorem you're not really making any assumption that
could be wrong, unless of course Lorentz invariance or quantum mechanics or
cluster decomposition is wrong, [\dots]
As long as you let it be the most general possible Lagrangian consistent with
the symmetries of the theory, you're simply writing down the most general
theory you could possibly write down. This point of view has been used in the
last fifteen years or so to justify the use of effective field
theories,[\dots].}
\end{quote}

At Chiral Dynamics 2009 in Bern, he replied to the question how to prove that:
\emph{I know of no proof, but I am sure it's true. That's why it's called a
  folk theorem.} It constitutes a lemma to what has since time immemorial been
known as ``Totalitarian Principle''\footnote{Often attributed to Gell-Mann,
  the origin is lost in words spoken long before 1956~\cite{Kragh}.} or
``Swiss Basic Law''~\cite{Ellis}: \emph{Everything not forbidden is
  compulsory.} That, in turn, is a corollary to the fundamental theorem
\emph{What-ever can happen will happen.}\footnote{In
  compliance with the Zeroth Theorem of the History of Science, the phrase's
  likely first appearance in print is a 1866 article by De
  Morgan~\cite{DeMorgan}, well before the famed Murphy was even born.}

If the ``Folk Theorem'' were all there is to it, an EFT would indeed be little
less than \emph{symmetries plus parametrisation of ignorance}\footnote{
  Originally a put-down, this dictum was embraced by EFT advocates early on;
  \cf~\emph{e.g.}~\cite{Daniel, Trento}} -- a Hail Mary to throw the kitchen
sink at any question. But an EFT involves much more: for an
\emph{efficient/effective} description at the scales \emph{of interest}, one
needs to identify the \emph{relevant} symmetries; the \emph{appropriate}
degrees of freedom; a \emph{workable} separation of scales; and a
\emph{consistent} scheme to bring order to the infinity of possible
contributions.

So an EFT actually offers ample opportunity to verify that anything that can
go wrong will go wrong. As discussed in sect.~\ref{sec:trust}, a series of
choices is based on assumptions, some obvious, some carefully stated, some
hidden, and at times even most dangerously hidden in plain sigh(t). Moreover,
the abstract theorem may hold, but since Physics is done by humans, Sociology
enters as discussed in sect.~\ref{sec:qqq}. We\footnote{Throughout, ``we''
  serves not as \emph{pluralis majestatis}, but as shorthand for the part of
  the nuclear community which is everybody but You, dear Reader.}  theorists
are often wrong, for example out of convenience (because doing the right thing
is hard); prejudice (because we know this to be wrong but it is not);
stubbornness (because we have always done it this way); lack of foresight
(thirty years later); or even sheer bad luck (why didn't anybody think of this
earlier?). What follows are a few instances which touched me; there are more
and better examples.

\begin{figure}[!h]
  \centerline{\includegraphics[width=0.45\linewidth]{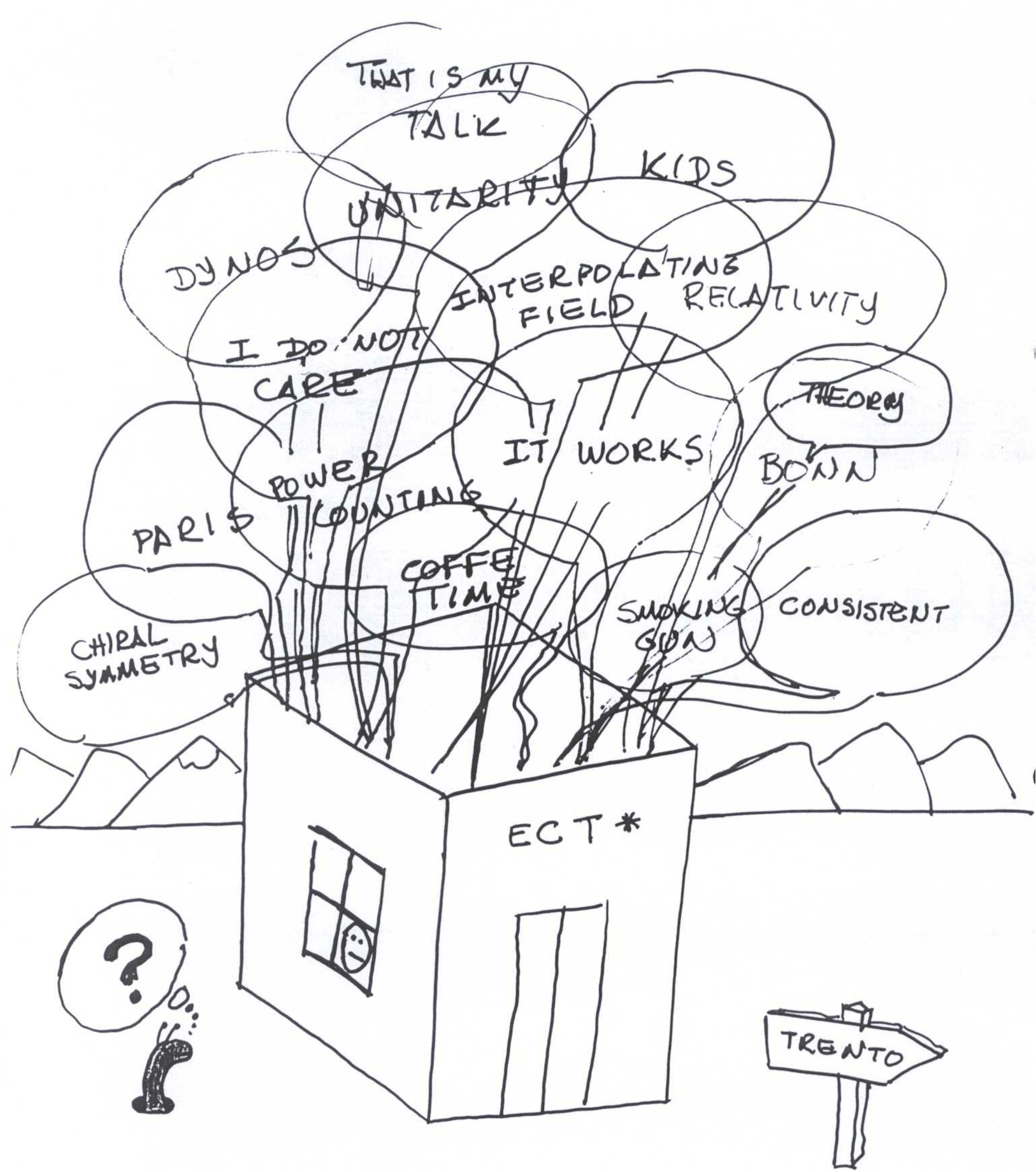}}
  \caption{Manoel Roberto Robilotta drew this cartoon the night before his
    talk at the 1999 \emph{Workshop on Nuclear
      Forces}~\cite{RobilottaCartoon}. I am grateful to Prof.~Robilotta for
    his generous permission to reprint a copy here.}
\label{fig:robilotta}       
\end{figure}

To illustrate the point that Physics is done by humans,
fig.~\ref{fig:robilotta} reproduces Manoel Roberto Robilotta's cartoon
capturing the spirit of the 1999 \emph{Workshop on Nuclear Forces}.  Notice
how familiar the topics sound even today. This was, I think, the first and
most epic clash between iconoclasts (commonly referred to as ``cockroaches''
there) and traditionalists (then called ``dinosaurs''), recalled also in
Ubirajara van Kolck's contribution to this issue~\cite{vanKolck:2021rqu}.  It
had been eight years since Weinberg's Nuclear Trifecta to whose central part
this issue is dedicated~\cite{Weinberg:1990rz, Weinberg:1991um,
  Weinberg:1992yk}\footnote{Reflecting another seismic change,
  \cite{Weinberg:1992yk} was Weinberg's first submission to the arXiv, which,
  incidentally, turns 30, too.}, but the nuclear community had taken little
notice, despite a seminal PhD thesis by an upstart about everything subsequent
EFT generations re-discovered~\cite{vanKolckPhD}.

\section{Input: Trust But Verify}
\label{sec:trust}

Let us first turn to some assumptions in Nuclear EFT -- from \emph{known knowns}
to \emph{unknown unknowns}~\cite{Rumsfeld}.

\subsection{Size Matters, Or: Issues With the Expansion Parameter}

Confronted with an infinite number of possible interactions, one must devise a
power counting scheme to order contributions and observables by their relative
importance in some small, dimensionless quantity
\begin{equation}
  \label{eq:Q}
  Q:=\frac{\mbox{typical low momenta } p_\mathrm{typ}}{\mbox{breakdown scale }
    \overline{\Lambda}_{\mathrm{EFT}}}<1\;\;.
\end{equation}
The numerator summarily depends on intrinsic low scales $p_\mathrm{typ}$ at
which the EFT is supposed to be applicable, including the relative momentum
$k$ between scattering particles, the masses of light particles ($\mpi$ for
the pion), and the scales associated with binding within the EFT's
reach. Around the breakdown scale $\overline{\Lambda}_{\mathrm{EFT}}$, new
dynamical degrees of freedom enter which are not explicitly accounted for by
the EFT but whose effects at short distances
$\lesssim1/\overline{\Lambda}_{\mathrm{EFT}}$ are simplified into Low-Energy
Coefficients (LECs). A breakdown scale is not a number, but a gradual corridor
of values over which an EFT becomes increasingly unreliable until it
eventually makes no sense at all.

The range is also limited from below. A Nuclear EFT assumes
$\ptyp\gg1\;\mathrm{eV}$ and dispatches atomic effects because that is not the
\emph{relevant} Physics at the requested scale. This implies one determines
parameters most efficiently around $\ptyp$, not at much smaller scales; see
the discussion of the fit corridor in sect.~\ref{sec:line}.

Take an observable $\obs$ whose first nonzero piece starts at order
$Q^{n_0}$. Let us label the $i$th-order contribution relative to leading order
(LO) as $Q^{i}\;\obs_i$, making the order of $Q$ explicit.  Discarding
logarithmic corrections for brevity's sake, it can hence be expanded as
\begin{equation}
  \label{eq:O}
  \obs=Q^{n_0}\left[\sum\limits_{i=0}^n Q^{i}\;\obs_i+\calO(Q^{n+1})\right]\;\;.
\end{equation}
One estimates the theory uncertainty from truncating the series at order
$O(Q^{n+n_0})$, or (Next-to-)${}^n$Leading Order (N$^n$LO) as being one order
higher, namely $Q^{n+1}$ relative to  LO and hence with an associated
truncation error of $Q^{n+1}$.

Consider the expansion of electro-dynamic effects. It proceeds in powers of
$\alpha=\frac{1}{137}$~\cite{Bethe}, so its higher-order corrections
contribute typically just $\lesssim1\%$. Compton scattering on a charge, for
example, starts at $\calO(\alpha^2)$, \ie~$n_0=2$, and the first correction is
at order $\alpha^3$, providing a correction to the LO result of
$\calO(\alpha^1)\approx1\%$.

But in Nuclear Physics, we are not blessed with an exceptionally small $Q$
even at typical low scales: $Q\approx0.4$ in \ChiEFT with a dynamical Delta
resonance degree of freedom, at best $\frac{1}{2}$ without Deltas, and
$\frac{1}{3}$ in \EFTNoPion are common numbers;
\cf~sect.~\ref{sec:Delta}. That makes estimating theory uncertainties even the
more imperative.  On top of that, there are quite a few options how to expand;
\cf~sect.~\ref{sec:sin}. Fortunately, one can check if expectation and outcome
match by carefully checking convergence patterns; see sect.~\ref{sec:line}.
When one invests that effort to dot the \i's, includes quite a few orders and
consequently faces quite a number of non-trivial interactions whose parameters
are usually determined well from the cornucopia of nuclear data, one is
rewarded by results with theory truncation errors which are both credible and
competitive with experimental errors -- and even overlap.

\subsection{It Ain't Natural, Or: Na\"ive Dimensional Analysis and Error Bars}
\label{sec:natural}

EFTs \emph{carry the seed of their own destruction}~\cite{Daniel}: At
$Q\gtrsim1$ ($\ptyp\gtrsim\LambdaEFT$), the Lagrangean may still perfectly
reasonably reflect the symmetries of the problem, but there is no power
counting.  Additional arguments must then justify why some terms are kept
while an infinity of others is dropped. Practicality is a good one, as is the
hope to ``model'' one's way to a more comprehensive understanding which
eventually may be cast into an EFT. Maybe one can rearrange the deck chairs to
find a converging result\dots
  
As $p_\mathrm{typ}\nearrow\overline{\Lambda}_{\mathrm{EFT}}$ ($Q\nearrow1$),
the demise of the power counting is of course not sudden but gradual. This
decrease of expected accuracy must be reflected in larger theory errors, for
example at higher energies, and accounted for both in data fits and in
comparing different EFTs; see \eg~\cite{JPhysG, JPhysG2, Melendez:2020ikd,
  Furnstahl:2021rfk}.

The expansion of $\obs$ in eq.~\eqref{eq:O} is based on a key assumption not
only of EFTs but of Physics in general: ``Weak Naturalness'' requires that
higher orders (namely so-called details) do generally not spoil the
perturbative series, \ie~$|\obs_{i}|>Q\;|\obs_{i+1}|$, with only ``a few''
exceptions; see also~\cite{tHooft:1979rat,Hammer:2019poc,vanKolckSaclay,
  Griesshammer:2020fwr}. When $Q\approx10^{-20}$ as in nuclear corrections
from Quantum Gravity at the Planck scale, ratios of
$|\obs_{i+1}|/|\obs_i|\approx10^{15}$ may appear prohibitively large, but
the contribution of the $(i+1)$st term is still suppressed by
$10^{-20+15}=10^{-5}$ against the $i$th term and hence provides a negligible
correction for all practical purposes. If, however, $Q\approx\frac{1}{4}$ as
in \ChiEFT, then ratios of $|\obs_{i+1}|/|\obs_i|\approx3$ or so are already
precarious. Thus, one often considers for example contributions from isovector
nucleonic magnetic moments, $\kappa_v\approx4.7\sim1/Q$ one order sooner,
avoiding relatively large but well-understood higher-order corrections; see
\eg~\cite{Griesshammer:2012we}.

Naturalness flows into another fundamental assumption: Higher-order terms can
reliably be estimated by Na\"ive Dimensional
Analysis~\cite{NDA,NDA2,Weinberg:1989dx, Georgi:1992dw,
  Griesshammer:2005ga}. Without these variants of Occam's Razor~\cite{Occam},
one cannot rule out alternative explanations via extraordinarily large
higher-order corrections. Since the dawn of the quantitative Scientific
Method, researchers have implicitly assumed that Nature is not malevolent\footnote{\emph{Raffiniert ist der Herrgott, aber boshaft ist er
nicht.}~\cite{Einstein}}. That makes the difference between Theory and Conspiracy Theory.

A comprehensive and quantitative theory of Weak Naturalness and Na\"ive
Dimensional Analysis has been emerging this past decade, based on checking
assumptions against outcomes using Bayesian statistics with reasonable
expectations clearly formulated as priors; see \eg~\cite{JPhysG,JPhysG2,
  Furnstahl:2021rfk, Melendez:2020ikd, Griesshammer:2020fwr} and references
therein. A cornerstone of any EFT is to actually provide quantitative
estimates of theory errors, rather than ``educated guesses'' based on ``years
of experience'', and we should have paid attention sooner;
\cf~sect.~\ref{sec:line}.

\subsection{No Freedom in the Degrees of Freedom?, Or: The Relevant Particle
  Content}

In Nuclear Theory, we may have largely found the ``right'' degrees of freedom
for efficient versions of the most general Lagrangean: ``pion-less EFT''
(\EFTNoPion) employs contact interactions between only nucleons (and external
probes) at very low momenta $p_\mathrm{typ}\ll\LambdaNoPion\approx\mpi$; and
Chiral EFT (\ChiEFT) adds pions and the Delta resonance in nuclear processes
at more generic nuclear scales $\ptyp\sim\mpi$; \cf~sect.~\ref{sec:Delta}. The
strange few-hadron sector has also been explored; see~\cite{Haidenbauer:2021wld}
in this issue and references therein.

The correlated two-pion state $f_0(500)$ could possibly be added as its own
degree of freedom. It has the quantum numbers of the QCD vacuum and a mass
somewhere around $([400\dots550]-[200\dots350]\ii)\;\MeV$ just around or below
$\LambdaChi$, according to the PDG's 2020 edition~\cite{Zyla:2020zbs}. Some
ideas about its significance for the $\N\N$ potential are emerging, especially
related to its r\^ole in the two-pion exchange; \cf~\cite{Donoghue:2006rg,
  Mishra:2021luw}.

It would be interesting to further explore its impact on few-nucleon systems,
and it would definitely be amusing if the pre-EFT
controversy\footnote{\emph{Come on, we can do without that sigma
    crap.}~\cite{Trento}} about the meson formerly known as $\sigma$ would be
revived. However, its enormous width and large mass suggest that it is just
as well captured by only very mildly energy/momentum-dependent LECs already in
\ChiEFT.

We have not yet found a path to formulate a clear separation of scales through
the jungle of GeV-scale meson and nucleon resonances. In such uncertain
territory, models and less-than-rigorous Ans\"atze provide crucial insight
into what degrees of freedom and symmetries may be appropriate and relevant --
if one optimistically assumes that we just have not yet found the right EFT
there.

As one moves to heavier nuclei, it is however no surprise that interactions
between ``free'' nucleons and pions become less \emph{efficient} ways to
describe the relevant Physics. Since the advent of the liquid-drop model, we
know that nucleons in heavy nuclei are not free but subject to some collective
motion. The bridge to descriptions which utilise more collective degrees of
freedoms, like shell-and-core or quasi-particles, is one we have now started
to explore with more confidence~\cite{Piarulli:2020mop,
  Drischler:2021kqh,Tews:2021bqc, Piarulli:2021ywl}; \cf~discussion in
sect.~\ref{sec:details}.

\subsection{The Delta Variant, Or: An Often-Overlooked Degree of Freedom}
\label{sec:Delta}

However, the $\Delta(1232)$ resonance still plays the r\^ole of the
understudy: used if unavoidable. Its resonance peak energy of about
$300\;\MeV\approx2\mpi$ above the nucleon mass, its width of
$\frac{\Gamma}{2}\approx70\;\MeV\approx\frac{\mpi}{2}$, and its rather sizeable
coupling to pions and photons, means its effects are manifest even at energies
$E\approx\Delta_M-\frac{\Gamma}{2}\lesssim200\;\MeV$.  The Delta channel opens
immediately with the pion threshold and has a dramatic energy dependence, as
well known from the textbook plots of cross sections in the first dozen $\MeV$
of pion-photoproduction and pion-nucleon scattering.  But even below that, at
$100\;\MeV$ or so, its impact is obvious in processes like few-nucleon Compton
scattering~\cite{Margaryan:2018opu} where energy and momentum of the probe are
actually identical. In that case, $E\sim\ptyp\sim\Delta_M$, so that the
breakdown of \ChiEFT without a dynamical Delta is at best set by the
Delta-nucleon mass splitting as
$\LambdaChi(\slashed{\Delta}) \lesssim\Delta_M\approx300\;\MeV$ -- but that
neglects that the large width makes it contribute even well before. Even in
that case, the expansion parameter
$Q=\frac{\ptyp\approx\mpi}{\LambdaChi (\slashed{\Delta})} \approx\frac{1}{2}$
would become uncomfortably large in processes like Compton scattering where
energy and momentum scales of external probes are identical. Whether results
actually converge, needs therefore close examination. Delta effects are at
times somewhat suppressed in isoscalar nuclei, but that is the exception to
the rule.

On the other hand, the breakdown scale $\LambdaChi\approx[700\dots1000]\;\MeV$
of \ChiEFT with dynamical Delta is consistent with the masses of the $\omega$
and $\rho$ as the next-lightest exchange mesons, and with the chiral symmetry
breaking scale; see~\cite{Pascalutsa:2002pi} for an oft-employed variant. At
$\ptyp\sim\mpi$, the expansion parameter is then about $\frac{1}{6}$ for pion
physics and about $0.4$ for the perturbative Delta. At energies
$\sim\Delta_M$, the Delta resonance dominates and constitutes LO,
$\pi\N\Delta$ interactions must be resummed, and the expansion parameter is
now about $0.4$ for pions as well. In either r\'egime, $Q$ is not very small,
but convergence appears quite reasonable, reaching $\lesssim\pm3\%$ around
$\mpi$ at \NXLO{4}. That is confirmed by a quantitative Bayesian analysis of
uncertainties which also bears in mind that both the power counting and
$Q$ itself changes with $\ptyp$~\cite{Melendez:2020ikd}.

Even at low energies, the Delta variant helps with Naturalness;
\cf~sect.~\ref{sec:natural}. Without it, some LECs like the $\pi\pi$N
coefficients $c_{2,3}$ are unnaturally large. Most of that strength is
resolved by resonance saturation as coming from the dynamical
Delta~\cite{Bernard:1996gq}. What remains of $c_{2,3}$ becomes
natural-sized. Even at low $p_\mathrm{typ}\sim\mpi$, that reduces the risk of
large corrections which are formally of higher order. So, the information from
additional degrees of freedom can actually improve predictions. So an improved
resolution, namely a ``more fundamental'' theory, can lead to an
increase of information, with fewer unknowns and fewer mysteries.

At $\ptyp\sim\mpi$, high-accuracy \ChiEFT interactions with a perturbative
Delta dramatically impact nuclear structure and nuclear
matter~\cite{Piarulli:2020mop, Drischler:2021kqh, Tews:2021bqc,
  Piarulli:2021ywl}. But for energies $\sim\Delta_M$, a typical scale if not
in finite nuclei, then in neutron stars, the Delta dominates over
pion-exchange and cannot be treated perturbatively. A theory without the Delta
is there computationally certainly more convenient than a coupled-channel
problem with it. Conceptually, \ChiEFT's standard should in the long run
be to include the Delta as a matter of course.

\subsection{These Are Not The Symmetries You Are Looking For, Or: The
  Importance of Conservation Laws}
\label{sec:symmetries}

Chiral symmetry, gauge and Lorentz invariance (often as perturbation in powers
of velocities; \cf~sect.~\ref{sec:variations}), and other symmetries are
certainly not over-constraining Nuclear EFTs. The cornucopia of non-trivial
high-accuracy agreements between theory and data makes it unlikely that we
have imposed an exact or approximate symmetry that is not there. Requiring
parity in weak interaction would be a counter-example. 

More tricky is the question whether we are missing symmetries. These would
show up as correlations between observables (and between parameters in the
Lagrangean) which appear accidental or fine-tuned.

Theorists do not like fine-tuning. They are much happier with an underlying
symmetry, even if approximate, to protect combinations of parameters from
deviating a lot after renormalisation. For example, chiral symmetry explains
why chiral-symmetry-breaking interactions are small and disappear as
$\mpi\to0$.

The most famed fine-tuning in Nuclear Physics is related to the anomalously
large S-wave scattering lengths $a$ and corresponding anomalously small
binding energies of few-N systems: the deuteron, triton and both Helium
isotopes. These need an intricate balance between attractive long-range and
repulsive short-range effects. Chiral symmetry alone does not explain this
fine-tuning.

Therefore, a few groups have proposed an expansion about a point where several
protective symmetries
coincide~\cite{Kievsky:2015dtk,Konig:2016utl,Konig:2016iny,
  Kolck:2017zzf,Kievsky:2018xsl,vanKolck:2019qea,Konig:2019xxk}. In the
unitarity limit $a\to\infty$, Nuclear Physics becomes scale-invariant at low
scales. The NN cross section saturates and no dimension-ful scale is
left. This may suggest that all nuclear binding energies must thus be infinite
or zero. But in \EFTNoPion, renormalisation via the Efimov effect breaks
continuous scale invariance down to a discrete one. That introduces one
dimension-ful $3\N$ scale, set for example by the binding energy of the
triton. When unitarity is imposed for both NN S-waves, Wigner's $SU(4)$
symmetry of arbitrary combined spin and isospin rotations is manifest as
well. So the unitarity limit is a point of increased symmetry, and Nature
appears to break it only by a small amount, $\frac{1}{\mpi
  a}\lesssim0.3$. This expansion reproduces well the ground state of $^4$He,
and that its first excitation is very close to breakup. There is even evidence
that the Coester correlation between binding energy per nucleon and nuclear
matter density can be explained, as well as the symmetry energy of nuclear
matter, its slope and compressibility.

Surprising here is not so much that the detailed value of the scattering
length becomes irrelevant in nuclei and nuclear matter -- after all, the
momentum scales are $\ptyp\gg\frac{1}{a}$. Surprising is that one might get
away with a theory of nuclear matter that does not know about pions if or
because $\ptyp<\mpi$. The typical binding momentum in heavy
nuclei is $\gamma_A\sim\sqrt{2MB_A/A}\to120\;\MeV\lesssim\mpi$ for
$B_A/A\approx8\;\MeV$, which is still less than the pion mass. Only more, and
more non-trivial, applications can shed light on this idea.

The goal here is not a detailed reproduction of Nuclear Physics, but a
conceptual understanding of its gross structure from its most important
symmetries. The proposal constitutes a paradigm shift away from emphasising
details of QCD and NN scattering, towards the importance of renormalised
scales in the 3N system and of approximate symmetries. It even speculates that
patterns like a Nuclear Chart are not unique to QCD but analogues emerge in
any many-body system with anomalously large scattering lengths, like clusters
of Rb atoms; \cf~\cite{Contessi:2021ydv}.  It also begs the obvious question
how this protective symmetry emerges for the EFT which includes pionic degrees
of freedom, and on the quark level. Finally, it sheds light on the fundamental
question how complex phenomena emerge from seemingly simple foundations, and
how simple patterns emerge in turn.

\subsection{The Original Sin\protect\footnote{Note Added in Proof:
    D.~R.~Phillips uses the same phrase to explore a slightly different
    perspective of the same issue~\cite{Phillips:2021yet}.}, Or:
  Power-Counting a Non-Perturbative LO}
\label{sec:sin}

Different choices how to count powers in the small dimension-less expansion
parameter $Q$ lead to vastly different physical situations, but they all need
to be consistent, not some \emph{ad hoc} prescription to be thrown overboard
once one encounters problems. Only self-consistent theories can be falsified
in observations. Inconsistent ones are wrong from the start\footnote{\emph{Das
    ist nicht nur nicht richtig; es ist nicht einmal falsch!}~\cite{Pauli}}.

When all interactions are perturbative, as in the mesonic and single-baryon
sectors, this amounts to little more than counting powers of $k$ and $\ptyp$
-- but not quite. One must still classify them as large or small relative to
another scale: $\mpi\gg m_e$ but $\mpi\ll M_\mathrm{Physicist}$. Usually, that
is understood, but dimension-ful quantities quickly trick one into paying less
attention to relevant high-momentum scales.

However, if there are shallow real or virtual bound states at scales
$\frac{1}{a}\lesssim\ptyp\ll\LambdaEFT$ in the EFT's range of validity, some
interactions must be treated non-perturbatively at leading order. In other
words, an infinite number of terms needs to be summed because bound particles
are never free -- they never do not interact with each other. Weinberg
ingeniously proposed to pragmatically power-count the potential, truncate it
at a given order, and then iterate that to create bound
states~\cite{Weinberg:1991um}; see also his more qualitative discussion
in~\cite{Weinberg:1990rz}. That appears to conflict with the fundamental EFT
tenet that the power counting only applies to physical (renormalised and/or
observable) quantities, but it was a convenient and clever way to enrol for
quick results the by-then well-matured technologies to solve Schr\"odinger and
Lippmann-Schwinger equations: plug a ready-made potential into an accepted
formalism.

Since then, \emph{who} employed this idea appears to have at times become more
important than \emph{whether} to employ it\footnote{\emph{When the President
    does it, that means that it is not illegal.}~\cite{Nixon}}. Especially in
the first decade of the third millennium, an exegetic, and at times even
hermeneutic, reading of the Sacred Texts~\cite{Weinberg:1990rz,
  Weinberg:1991um,Weinberg:1992yk} attempted to extricate more than a plain,
constitutionalist interpretation allows. While research is often inspired
by particular phrases we encounter, scientists can fortunately claim to be
agnostic about an author's intentions or reputation. What eventually counts
(or should count) is logical deduction and reproducible observation.

Consider a general argument much older than
ref.~\cite{Griesshammer:2020fwr}. Denote the NN scattering amplitude
$T_\mathrm{NN}$ by an ellipse and the interaction kernel $K_{2\N}$ by a
rectangle. For two nucleons, the kernel is the two-nucleon potential
$V_\mathrm{NN}$ of strong interactions. The semi-graphical representation of
the well-known LO Lippmann-Schwinger integral equation is:
\begin{equation}
  \label{eq:consistency}
  \begin{split}
    \includegraphics[clip=,width=0.6\linewidth]{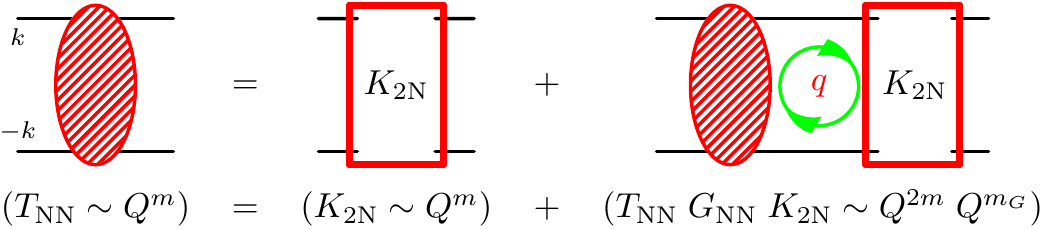}
  \end{split}
\end{equation}
where $\qv$ is the relative momentum of the nucleons in the intermediate
state, $\kv$ is the scattering momentum, and nucleons are close to their
non-relativistic mass-shell, $E\sim\frac{k^2}{M}\sim Q^2$ (potential r\'egime;
see \emph{e.g.}~\cite{Beneke:1997zp, Griesshammer:1997wz}).  The intermediate
$\mathrm{NN}$ state of free two-nucleon propagation is described by a
propagator (free Green's function of the NN system) and an integration. Let us
say this operator scales with some power of $Q$,
\begin{equation}
  \label{eq:GNN}
  G_{\mathrm{NN}}:=\int\deintdim{3}{q}\frac{1}{\kv^2-\qv^2}\sim Q^{m_G}\;\;.
\end{equation}

Only a nonperturbative solution, namely summing at least some subset of
interactions an infinite number of times, keeps the particles always close and
hence creates a bound state.  Therefore, all terms, including the interaction,
must be of the same order. Without that, one term could be treated as
perturbation of the others.  For example, if the driving term (in the middle)
were of higher order, we would see bound states but no scattering states. If
the homogene\"ity (last term) were of higher order, one would find the Born approximation.

That resummation is not just a good idea when shallow bound states exist, but
is compulsory, imposes consistency conditions on both interaction and
amplitude: they both must count the same,
$T_\mathrm{NN}\sim K_{2\N}\sim Q^m$, and the last term in
eq.~\eqref{eq:consistency} imposes they must count inverse to $G_\mathrm{NN}$,
\ie~$m=-m_G$. The integral in $G_{\N\N}$ is dominated by the parts with large
integrand, namely when typical scales of loop momenta are the external, low
momenta: $k\sim q\sim\ptyp\sim Q$. That is also a fundamental tenet of Na\"ive
Dimensional Analysis justified in the threshold expansion
formalism~\cite{Beneke:1997zp, Griesshammer:1997wz}.  Therefore, the scaling
is fixed as $G_\mathrm{NN}\sim Q^{3-2}=Q^1\sim Q^{-m}$ and the mere existence
of a shallow real or virtual bound state mandates
$T_{\mathrm{NN}}\sim K_{2\N}\sim Q^{-1}$.

Remember that $G_\mathrm{NN}$ describes the propagation of two non-interacting
nucleons between two interactions. It is agnostic about the kernel
$K_{2\N}$. Ultimately, binding must be explained by the interaction as
the origin of an intimate correlation of nucleons, and not by the free-nucleon
propagator.


This reasoning has several intriguing aspects. It is simple. It
only relies on the qualitative feature of the \emph{existence} of an
anomalously shallow bound state, not on any particular value of $a$.  It does
not reveal \emph{which} terms constitute the LO kernel or how the shallow
scale emerges from it; only how those terms must be power-counted. It thus
equally well applies to any systems with shallow bound states, including
halo-EFT, \EFTNoPion and Non-Relativistic QED/QCD. It is consistent at LO and
permits corrections to be treated in strict perturbation theory; see
sect.~\ref{sec:perturbation} below. It imposes a power-counting \emph{from} an
observable, namely the scattering amplitude $T_\mathrm{NN}$, \emph{via} a free
propagator $G_\mathrm{NN}$ \emph{onto} the non-observable $K_{2\N}$, not
\emph{vice versa}.

It also leads to a surprising take on the one-pion exchange, as it appears to
scale like
$\frac{(\vec{\sigma}_1\cdot\vec{q})(\vec{\sigma}_2\cdot\qv)}{\qv^2+\mpi^2}
\stackrel{??}{\sim} Q^0$ if one counts only explicit low-momentum scales, but
must be of order $Q^{-1}$ if its iteration is required.

Consequently, a choice of \ChiEFT{}s exist with the same symmetries and
degrees of freedom but different power countings, corresponding to different
worlds. In the ``KSW'' version, the system is at LO ($Q^{-1}$) bound by
contact interactions only, like in \EFTNoPion, and the one-pion exchange
scales indeed as $Q^0$ to enter at NLO. Its analytic results in the
$\mathrm{NN}$ system pass every test of
self-consistency~\cite{Kaplan:1998tg,Kaplan:1998we, Fleming:1999ee};
\cf~sect.~\ref{sec:elders}.

In the most popular version of \ChiEFT, the one-pion exchange is taken to
enter at leading order~\cite{Weinberg:1990rz, Weinberg:1991um,
  Weinberg:1992yk} and therefore \emph{must} scale as
$\frac{(\vec{\sigma}_1\cdot\vec{q})(\vec{\sigma}_2\cdot\qv)}{\qv^2+\mpi^2}\sim
Q^{-1}$. One cannot just count momenta.

There are many other versions, all consistent, but all describing different
worlds.  One has no shallow scales ($\frac{1}{a}\sim\LambdaEFT$), one-pion
exchange
$\frac{(\vec{\sigma}_1\cdot\vec{q})(\vec{\sigma}_2\cdot\qv)}{\qv^2+\mpi^2}\sim
Q^0$ is perturbative (Born approximation) and only detailed knowledge of QCD
explains nuclear states. We quickly throw that one away because Nature has
bound states in low partial waves\footnote{Note Added In Proof: See footnote 2
  in D.~R.~Phillips' contribution to this issue~\cite{Phillips:2021yet}.}. But
its power-counting is appropriate for higher ones~\cite{Kaiser:1997mw}.

Most likely, the real world is reproduced at LO by mix-and-match: contact
interactions plus one-pion exchange in the \wave{3}{SD}{1} and some other low
partial waves, KSW in \wave{1}{S}{0} and others, and perturbative in higher
partial waves $l\gtrsim3$ or so~\cite{Beane:2001bc,
  Barford:2002je,Birse:2005um,Birse:2009my};
\cf~sect.~\ref{sec:nonperturbative}.

So what makes one decide whether resumming one-pion exchange (OPE) at LO is
mandatory or discretionary?  Its scale appears in \ChiEFT to be set by
$\Lambda_\mathrm{NN}=\frac{16\pi f_\pi^2}{g_A^2
  M}\approx300\;\MeV$~\cite{Kaplan:1998tg,Kaplan:1998we,Barford:2002je}. That
lies right between the typical low scale $\mpi$ and the expected breakdown
scale $\LambdaChi$. This scale is dynamic, dictated by interactions, and thus
most naturally accommodated in the kernel/potential $K_{2\N}$. Below it, pions
are higher-order effects and hence perturbative (KSW); above, they are LO and
hence nonperturbative. However, that does not explain why shallow bound states
exist. Chiral extrapolations show that the QCD parameters are fine tuned to
produce large scattering lengths at the physical pion mass;
see~\cite{Kirscher:2015tka} and references therein. Even small variations in
$\mpi$ bring one quickly to a world with $a\lesssim\frac{1}{\mpi}$. But
$\Lambda_\mathrm{NN}$ is largely constant in $\mpi$ since it only involves
$g_A$, $f_\pi$ and $M$, none of which change dramatically with $\mpi$. For
$\ptyp\gtrsim\Lambda_\mathrm{NN}$, one may thus be forced to resum the one
pion exchange even without shallow bound states and even where there is no
fine tuning, namely well off $\mpi\approx140\;\MeV$.

Weinberg's pragmatic proposal is widely interpreted as counting powers of
$\ptyp\sim q$ only, with $K_{2\N}\sim q^0\sim Q^{m=0}$ as LO. Since
there is a shallow bound state, eq.~\eqref{eq:consistency} mandates then
$G_\mathrm{NN}\sim Q^0$, and therefore $q\sim Q^0$. That leads to the
contradiction that all powers of $q$ actually count the same, $q^n\sim Q^0$ --
unless one resorts to fine-tuning in $G_\mathrm{NN}$. One could try to count
powers of $q$ and of $(k^2-q^2)$ differently in eq.~\eqref{eq:GNN}. But that
contradicts that typical scales of loop momenta are the external, low momenta
$\ptyp$.  One could also try to fine-tune otherwise independent
contributions. For example, re-interpret $G_\mathrm{NN}$ as doing more than
just propagating two free nucleons. Since eq.~\eqref{eq:consistency} is
form-invariant under
\begin{equation}
  G_\mathrm{NN}\to Q^{-1} \;G_\mathrm{NN}\sim Q^0\;\;,\;\;T_\mathrm{NN}\to
  Q\;T_\mathrm{NN}\sim Q^0\;\;,\;\;K_{2\N}\to Q\;K_{2\N}\sim Q^0\;\;,
\end{equation}
this shifts the fine-tuning burden from the kernel $K_{2\N}$ to the
two-nucleon propagator $G_\mathrm{NN}$ and turns the free propagator into a
quasi-correlated one -- except that there are no interactions in it. I fear
this leads to a contradiction: The propagator is both propagating without
interactions and knows it is inside a nucleus, subject to binding forces and
correlations with other nucleons -- it \emph{is born free but everywhere is in
  chains}~\cite{Rousseau}. On top of that, none of this explains how to
proceed with the power counting at higher orders. Often, Weinberg's pragmatic
proposal is advocated as advantageous because one only needs to count powers
of $\ptyp$ in the kernel $K$, but what about $G_\mathrm{NN}$? Does its
fine-tuning persist at higher orders, or does it also contain contributions
which scale naturally?  As to be discussed in sect.~\ref{sec:toymodel}, the
idea already fails at NLO. Without the alleged fine tuning, on the other
hand, it counts reproducibly as $Q^1$.
 
Finally, one can extend this construction to $n$-nucleon systems with a kernel
$K_{n\N}$ by induction from the scaling of the amplitude $T_{(n-1)\N}$ of the
anomalous $(n-1)$-particle subsystem with kernel $K_{(n-1)\N}$:
\begin{equation}
  T_{n\mathrm{N}}\sim K_{n\mathrm{N}}\sim Q^{1-n}\;\;.
\end{equation}
Remember that $K_{n\mathrm{N}}$ is \emph{not necessarily} what is often called
a $n$-nucleon interaction, but rather a kernel which involves interactions of
possibly up to $n$ nucleons. It could also only contain $\N\N$
interactions. The construction is agnostic about which interactions are needed
-- it only addresses how those which are needed, are to be power-counted. In
\ChiEFT, for example, the leading-order kernel $K_{3\mathrm{N}}$ between $3$
nucleons appears to contain only $\N\N$ interactions, and no $3\N$
interactions~\cite{Nogga:2005hy}. In \EFTNoPion, however, it is well known
that a $3\N$ interaction is needed at LO to stabilise the system against
collapse, in addition to an iterated $\N\N$ interaction. So there is no
contradiction between the $3\N$ interactions of these two theories because
their information content is different. What exactly constitutes the LO kernel
depends on the particles, symmetries, resolution scales~\etc~of the theory.

\subsection{It Works Until It Doesn't, Or: Power Counting for Uncorrelated Nucleons}
\label{sec:works}

The preceding argument also defines a correlated few-nucleon propagator: Even
the $m$th iteration of the $n$-nucleon interaction counts the same, namely
$K_{n\N}\sim K_{n\N}(G_{n\N}K_{n\N})^m\sim T_{n\N}\sim Q^{1-n}$. It is this
rescattering series which turns the free Green's operator into the correlated
propagator of nucleons inside a nucleus,
\begin{equation}
  G_{n\N}\;\to\;G_{n\N}^\mathrm{correlated}=
  G_{n\N}\sum\limits_{i=0}^\infty \left(K_{n\N}G_{n\N}\right)^i
  =G_{n\N}\left[\one-K_{n\N}G_{n\N}\right]^{-1}\;\;.
\end{equation}
One can now understand the transition from correlated to
quasi-free/uncorrelated propagation of nucleons inside nuclei, as formalised
in Threshold Expansion~\cite{Beneke:1997zp, Griesshammer:1997wz}. The question
becomes: At which scales is the resummation of the geometric series in the
equation above mandatory, and at which is it merely optional?

A key assumption above was that the scattering particles are not far off their
mass shell, $k^2\sim q^2\sim ME$. Since this implies that momenta are much
bigger than kinetic energies, $k\sim q\gg E$, energy flow in interactions can
be treated perturbatively, which leads to a kernel $K_{2\N}$ which
describes an instantaneous, energy-independent potential $V_\mathrm{NN}$
between two nucleons at LO -- like in one-pion exchange.  What happens when an
external probe, say a photon, pumps more energy $\omega$ into the system?  The
relative importance of terms, and hence the power counting, changes with
energy because the Physics that is relevant changes qualitatively.  First,
adding or subtracting external energy modifies the propagator in $G_{\N\N}$ to
\begin{equation}
  \label{eq:external}
  \frac{1}{k^2-q^2}\;\to\;\frac{1}{k^2\pm M\omega-q^2}
\end{equation}
For $\omega\sim E\sim\frac{k^2,q^2}{M}\ll k,q$, that little more of energy
does not knock nucleons significantly off their mass shell. The scale of the
$q$-integration in eq.~\eqref{eq:GNN} is still set by $k$,
\ie~$G_{n\N}\sim Q^1$ and $T\sim Q^{-1}$ as before. Nucleons interact for a
long time $\sim\frac{1}{\omega}\gg\frac{1}{k}$, until the uncertainty
principle makes them radiate the excess energy, \eg~as another photon. In that
time between energy absorption and emission, such rescattering maintains the
the correlated nuclear state $G_{n\N}^\mathrm{correlated}$. This is precisely
what is needed in Compton scattering at $\omega\ll\mpi$ to maintain the
Thomson limit as low-energy theorem on the nucleus as a whole. One can show
that it is fulfilled exactly at LO, with zero higher-order corrections;
see~\cite{Griesshammer:2012we} and references therein.

On the other hand, nuclear coherence breaks at higher energies
$\omega\sim k\gg E\sim\frac{k^2}{M}$ (but of course still
$\omega\ll\LambdaEFT$). Such scales are relevant in Compton
scattering~\cite{Griesshammer:2012we}, and in particular in pion
photo-production~\cite{Beane:1997iv} and pion-nucleus
scattering~\cite{Weinberg:1992yk}, where the pion's absorption by the nucleus
adds an energy of at least $\omega\ge\mpi$. Intuitively, the
intermediate-nucleon state is in that case far off-shell and the time-scale
$\frac{1}{\omega}\lesssim\frac{1}{k\sim\mpi}$ between absorbing and emitting
the large excess energy is too short for much rescattering. Instead, nucleons
propagate largely un-correlated, namely quasi-free via $G_{n\N}$. Rescattering
is demoted to a higher-order effect. Indeed, expand eq.~\eqref{eq:external} as
\begin{equation}
  \frac{1}{k^2\pm M\omega-q^2}
  \stackrel{M\omega\gg k^2}{\longrightarrow}\frac{1}{\pm M\omega-q^2}
  \left[1+\calO\left(\frac{k^2}{M\omega}\right)\right]
\end{equation}
reveals that the information about binding (momentum $k$) is lost at LO and
the momentum scale in the integration is set by the external probe as
$q^2\sim M\omega$. The binding scale $k\sim\frac{1}{a}$ has disappeared, and
there is no need to iterate an instantaneous kernel $K_{n\N}$ for
$T_{n\N}$. Instead, the energy $\sim\omega$ and momentum $q$ of the
intermediate state are of the same size, which means retardation becomes
important. The interaction between nucleons is not any more instantaneous, and
$K_{n\N}\sim Q^0$.

Weinberg's pion-deuteron scattering used that the pion transfers
$\omega\ge\mpi$ on the nucleus and allows for at most one instantaneous
charged-pion exchange between nucleons before the pion is radiated off
again~\cite{Weinberg:1992yk}. He intuitively identified a two-nucleon
irreducible part which was made coherent by iteration to produce the nucleus,
and then sandwiched irreducible chiral interactions with external probes once
between the resulting wave functions, at energies so high that intermediate
states had no coherence.

\subsection{Child's Play, Or: Why Getting Counting Right Beyond LO Is
  Important}
\label{sec:toymodel}

Toy models are helpful idealisations because lessons in simpler settings
inform the search for solutions of complicated issues. While they can validate
hypotheses only for the model studied, and not in the general case, they can
invalidate general hypotheses by counter-example. In Nuclear Physics, we have
a toy model which even provides useful Physics insight: \EFTNoPion; \cf~the
ideas about the unitarity limit in sect.~\ref{sec:symmetries}.

\EFTNoPion also illustrates that power-counting is not \emph{just Mathematics,
  no Physics}~\cite{Trento}: One cannot cure deficits by just going to
sufficiently high orders until the actually-lower-order LECs are all
included. An illustration might be useful. In Weinberg's pragmatic proposal,
the order at which each term in S-wave \EFTNoPion contributes should just
follow the number of momenta, namely $p^{2n}\,C_{2n}\sim Q^{2n}$ enters at
\NXLO{2n}. At LO or alleged $\calO(Q^0)$, the scattering lengths suffice as
sole input. No correction enters at what is believed to be $\calO(Q^1)$, and
$p^2C_2$ enters only at putative $\calO(Q^2)$ or \NXLO{2}, determined by one
new datum. Assuming for concreteness $Q\approx\frac{1}{3}$, \NXLO{2} would
only provide a correction of $Q^2\approx\pm10\%$; the next term, allegedly
\NXLO{4}, just $Q^4\approx1\%$.

Actually, one can show analytically that the contact interactions of
\EFTNoPion scale as $p^{2n}\,C_{2n}\sim Q^{n-1}$. LO is $Q^{-1}$ as expected,
and corrections enter at \NXLO{n}, namely much earlier than anticipated in the
pragmatic proposal. For example, $p^2C_2$ is NLO and contributes
$Q^1\approx\pm30\%$. That re-ordering ricochets across orders. In a little
twist, the renormalisation group flow of the $n$th term allows for a new
parameter to be determined from data only at
\NXLO{2n-1}~\cite{Fleming:1999ee}. So, every subsequent fit parameter $C_{2n}$
enters one order earlier than conjectured, and its effect is
$\frac{1}{Q}\approx3$ times stronger than when simplistically counting
momenta.


In conclusion, an incorrect power counting scheme to classify interactions
makes one work not to an accuracy one had hoped for, but bears the very real
danger of doing worse. The optimist is then frustrated that Bayesian
uncertainty quantification~\cite{JPhysG, JPhysG2, Furnstahl:2021rfk,
  Melendez:2020ikd} reveals the corrections from one alleged order to another
to be quite frequently larger than anticipated, until one is finally forced to
acknowledge that something must be wrong.  One may then ask if fighting the
obvious at all cost is worth years of trouble.


\section{Development: Humanity Amongst Physicists}
\label{sec:qqq}

Let us therefore talk about pride and prejudice.

\subsection{The Whole Is Greater Than The Sum Of Its Parts, Or: Testing
  Consistency in Non-Perturbative LO}
\label{sec:nonperturbative}

So developing a consistent power counting is a trifle more involved than just
counting powers of momenta and asserting that that is all there is to
it. Instead, consistency needs to be checked at each order. It is no wonder
that once Particle Physicists had figured out the basics of QCD and a possible
path to Confinement, some moved on to perturbative Physics at higher energies
and thus intellectually less taxing issues than nonperturbative power
counting, emergence and complexity\footnote{I write as someone raised as
  Particle Physicist, inoculated early with a contempt for the messiness of
  Nuclear Physics.}.

After a series of critiques of Weinberg's approach based on analysing diagrams
or classes of diagrams which need to be resummed starting
with~\cite{Kaplan:1996xu}, Beane \etal~\cite{Beane:2001bc} and Nogga
\etal~\cite{Nogga:2005hy} employed fully non-perturbative arguments to
demonstrate that Weinberg's pragmatic proposal is fundamentally and
irredeemably flawed not just because the argument why LO should be resummed is
not consistent.

Beane \etal~\cite{Beane:2001bc} showed that in Weinberg's pragmatic proposal,
the $\mpi$-dependence of contact interactions between two nucleons does not
match that of the cutoff-dependencies they are to cure, for example in the
\wave{1}{S}{0} wave. But extrapolations to physical pion masses used in
lattice-QCD rely on the correct $\mpi$-dependence of LECs. To those
un-interested in relating \ChiEFT to QCD, that appears irrelevant in the real
world where $\mpi\approx140\;\MeV$ is fixed, until one realises that chiral
minimal substitution generates interactions from such dependencies. Chiral
symmetry dictates that a $\mpi^2$-dependent $\N\N$ interaction creates at the
next chiral order an interaction between two pions and two nucleons which has
the same strength:
$D_2\mpi^2\N\N\stackrel{\chi\text{sym}}{\longrightarrow} D_2\mpi^2\pi\pi\N\N$
\etc~This child inherits a power counting and strength from its parent.

Just as concerning is the finding by Nogga \etal~\cite{Nogga:2005hy} that the
pragmatic proposal cannot cure strong cutoff-dependencies in some of the
important attractive $\N\N$ partial waves. The argument is actually quite
intuitive~\cite{Barford:2002je,Birse:2005um,Birse:2009my}. The tensor part of
one-pion exchange contributes at LO. When it is attractive $\propto r^{-3}$ at
small distances, the wave function collapses into the origin, meaning one is
sensitive to details of short-distance Physics. To avoid that, one must add a
repulsive LEC with one parameter determined by data. That is the familiar
scenario in the \wave{3}{SD}{1} wave. Whether that happens in other waves as
well depends on the height of the repulsive centrifugal barrier
$\propto l(l+1)r^{-2}$ of orbital angular momentum $l$, relative to the
scattering energy and the strength of the tensor interaction, at distances
$r\gtrsim\LambdaChi$. Peripheral waves remain unaffected because the
centrifugal barrier is too strong~\cite{Kaiser:1997mw}, but low partial waves,
most notably \wave{3}{P}{0} and \wave{3}{PF}{2}, need LECs even at LO. Since
these are momentum-dependent like $k^{2,4,6,8}$, they would only enter at
\NXLO{2,4,6,8} in Weinberg's pragmatic proposal, but are actually needed at LO
to prevent collapse. The effect of these momentum-dependent, stabilising
interactions is large, and again ricochets across orders like in \EFTNoPion;
\cf~sect.~\ref{sec:toymodel}. On top of that, minimal substitution turns them
again into additional interactions with external probes at leading and
subsequent orders, \eg~in photo-nuclear reactions via LECs for $\gamma\N\N$,
$\gamma\pi\N\N$ \etc~Likewise, gauge and chirally invariant few-nucleon LECs
are re-ordered~\cite{Valderrama:2014vra, Hoferichter:2015ipa}.

As sociological footnote, it took EFT practitioners more than a decade, and
several since the discovery of the tensor force, to numerically test if it is
properly renormalised by momentum-independent LECs. What Nogga
\etal~\cite{Nogga:2005hy} did is so endearing because it is straightforward,
almost trivial -- in hindsight -- and thus begs the question why it had not
been done before.

The mis-classification in Weinberg's pragmatic proposal translates thus into
mis-estimates of coupling strengths, and thus of the accuracy to which
single-nucleon observables like the $\pi\N$ scattering lengths can be
extracted from few-nucleon data before few-nucleon LECs like a $\pi\pi\N\N$
term enter. Under-estimating their importance leads to a false sense of
accuracy. Weinberg's pragmatic proposal may predict that one can extract some
one-nucleon observables from few-nucleon data with an accuracy of $\pm5\%$,
when it is actually only $\pm20\%$ -- unless data from different nuclei allow
one to extract the new few-nucleon LEC or these can be determined in
lattice-QCD computations like in~\cite{Beane:2015yha}.

J.~de Vries and collaborators recently demonstrated that the consequences are
beyond ivory tower theatralics; see \eg~\cite{Cirigliano:2018hja,
  Cirigliano:2020dmx}. Building on~\cite{Valderrama:2014vra}, they showed that
in neutrinoless double-$\beta$ decay, a short-range, lepton-number-violating
interaction $\text{nn}\to\text{ppee}$ between two neutrons already enters at
LO, and not at \NXLO{2} as Weinberg's pragmatic proposal would have
it. Therefore, one cannot make even LO predictions of $0\nu\beta\beta$ matrix
elements in nuclei without at least estimating the strength of that
interaction. Fortunately, they were also able to calculate the corresponding
LEC. That is of course indispensable for any interpretation. Computations of
the pertinent matrix elements are now being redone by several
groups. Likewise, few-nucleon interactions with unknown coefficients enter at
NLO in the direct detection of Dark Matter via nuclei and in the search for
Electric Dipole Moments. Notice that these are all processes in which planning
and analysis of multi-million-dollar experimental efforts to look for
beyond-the-Standard-Model Physics rely on theory predicting, not post-dicting,
effects.

\subsection{There is Always a Well-Known Solution to Every Human Problem --
  Neat, Plausible, and Wrong~\protect\cite{Mencken}}
\label{sec:PC}

So, ordering interactions in \ChiEFT is not as simple a prescription as adding
and subtracting powers of momenta. It is a set of operational instructions:
Include at each order only those interactions needed to renormalise the
problem or with coefficients which Na\"ive Dimensional Analysis predicts at
that order.

Fortunately, a small band of brave theorists has taken on the ungrateful and
gruelling task to tirelessly turn such abstract rules into tables for the rest
of us what term needs to be added at what order~\cite{Birse:2005um,
  Birse:2009my, Valderrama:2009ei, Valderrama:2011mv, Valderrama:2019lhj,
  Long:2011xw, Long:2012ve}; see~\cite{Phillips:2013fia} for an even-handed
review. The dispute which one is correct is not yet completely settled
and in affectionate circles known as \emph{Power Counting Wars}\footnote{I am
  sure there is a dissertation about pop culture references in the oral
  scientific discourse. Please let me know.}.  A lack of universally accepted
analytic solutions obfuscates the relation between cutoff-independence,
convergence pattern and numerics, so Bayesian analysis and
renormalisation-group consistency checks are reasonable tools to convince the
community~\cite{JPhysG, JPhysG2, Furnstahl:2021rfk, Melendez:2020ikd,
  Griesshammer:2020fwr}.

Whether we will listen to and implement the outcome, will determine the fate
of \ChiEFT as either another set of models which have at least (largely) the
correct symmetries but to which parameters are added as needed to match data
-- or as a comprehensive and consistent theory of nuclear phenomena;
\cf~sect.~\ref{sec:details}.

That does not mean we on the sidelines need to wait with bated breath until
the appropriate power counting is established and a consistent set of
interactions between pions, nucleons, Deltas and external probes is available,
with tested and widely-accepted prescriptions to estimate residual theory
errors, to assess residual renorma\-lisation-group (cutoff) dependence, and to
establish World Peace. Instead, work progresses now in parallel to update few-
and many-body codes with new chiral interactions, all of which will eventually
contribute at some order, and with routines to assess uncertainties.

It is indisputable, however, that Weinberg's pragmatic proposal is 
not the way forward because
ultimately, and leaving all arguments of the preceding sub-section about its
self-consistency aside, it pays too much attention to terms that do not matter
that much, and not enough attention to terms that matter more than one might
have thought. It makes us work both more and less than needed, and it lulls us
into a false sense of accuracy. A brilliant idea got us started on the right
track, and it turned out to be pioneering but wrong after we learned from it
how to think for ourselves.

While the final verdict on what interactions to add at which order is still
pending, a number of crucial features are already decided, including that
there are more LECs in the attractive triplet-P waves. Today's potentials,
flawed as they are, already show interesting trends.
Important lessons are already learned from consistent power counting in
not-so-light nuclei; see \eg~most recently~\cite{Yang:2020pgi},
and~\cite{Furnstahl:2021rfk} in this issue.

\subsection{Fit To Shrink, Or: How Much Can We Learn From $\N\N$ Data?}

With well over $6,000$ $\N\N$ scattering data, the temptation is big to turn
into ``chisquare afficiados''~\cite{Trento}, trying to reproduce relatively
narrow but cornucopious information extremely well.  But if we really need a
$\chi^2$ close to $1$ in NN phase shifts and binding energies, plus maybe in
3N and 4N, to achieve even just several percent of accuracy for ground and
excited energies in nuclei -- arguably the observables least sensitive to how
well one's wave function actually captures reality -- then that may indicate
another fine-tuning: very many orders in \ChiEFT conspiring, overlapping,
cancelling and enhancing each other. The idea is not far fetched that the
power-counting/relative importance changes in not-so-few and many-nucleon
systems to de-emphasise one-pion exchange. After all, the precise values of
the scattering lengths play certainly no substantial r\^ole in nuclear matter;
\cf~sect.~\ref{sec:symmetries}. Indeed, combinatoric arguments have recently
been employed to advocate for greater relevance of three- and more-nucleon
interactions~\cite{Yang:2021vxa}.

In a way, the strange sector of \ChiEFT is more fortunate: Data exists but is
rare and not of the kind of alleged extraordinary quality which distracts from
the core mission to explain, rather than to fit. It also encourages one to be
much more ingenious to determine parameters; see, for example,
\cite{Haidenbauer:2021wld} in this issue and references therein. Sometimes,
less is more, and too much can be a curse.

\subsection{The World Is Not All That Is $\N\N$, Or: The Value of External Probes}

Of course, Nuclear Physics is more than $\N\N$ and few-$\N$ bound and
scattering states. One can learn just as much, and often complementing,
information from external probes, breakup, fusion, \etc~For example, pion
scattering~\cite{Weinberg:1992yk}, pion photoproduction~\cite{Beane:1997iv}
and Compton scattering~\cite{Griesshammer:2012we} on light nuclei all test the
charged pion-exchange contribution to nuclear binding, and thus chiral
symmetry.

\subsection{Do Not Listen To Your Elders, Or: Calculating Higher Orders Made Simpler}
\label{sec:perturbation}

Following Weinberg's example~\cite{Weinberg:1991um, Weinberg:1992yk},
observables beyond LO have traditionally been found by ``partial
resummation'': Power-count the strong interactions in few-nucleon systems
(usually the wrong way; \cf~discussions above); truncate at a desired order;
and then iterate by inserting it into eq.~\eqref{eq:consistency}. Likewise,
power-count interactions with perturbative external probes, and then sandwich
between the wave functions derived from the partially resummed few-nucleon
interactions. Long ago, we actually followed Weinberg's ``hybrid
approach''~\cite{Weinberg:1992yk} to use any high-precision wave function,
chiral or not. However, this leads to a mismatch of unphysical high-momentum
components and exacerbates theory errors; see also sect.~\ref{sec:variations}.
Fortunately, the advent of high-quality chiral potentials allowed us to move
on.

For the EFT power counting to make sense, higher-order corrections must be
ever-smaller. So including them in ``strict perturbation'' must be
allowed. The geometric series provides a nice example. For $|x|\ll1$, the
resummed and expanded-but-truncated versions must agree within truncation
errors:
\begin{equation}
  \frac{1}{1-x}-\left(1+x+x^2\right)=\calO(x^3)\ll1\mbox{ for } |x|\ll1\;\;.
\end{equation}
In general, if the resummed and strictly perturbative results differ
significantly, then corrections are obviously not actually small and one is
faced with fine-tuning.  Whether to resum at higher order or not should
therefore not be a matter of principle, but of choice and convenience;
\cf~kinematics example in sect.~\ref{sec:line}.  But strict perturbation also
avoids a number of problems.

First and pragmatically, including complicated interactions in strict
perturbation often avoids solving differential or integral equations and leads
to simpler, more stable numerics. It is thus not an uncommon trick when 3N and
4N interactions are added in nuclei and nuclear matter; see
\eg~\cite{Witala:2021ufh} in this issue.

Second, iterations usually generate spurious deeply bound states. While by
definition outside the EFT's range of applicability, these are often
precariously close and infect observables even at $\ptyp<\LambdaEFT$. With
higher order or higher cutoff $\Lambda$, their number proliferates and they
become more problematic~\cite{improve3body,Vanasse:2013sda}. Take the resummed
Effective Range Expansion:
\begin{equation}
  \label{eq:ERE}
  -\frac{4\pi}{M}\;\frac{1}{\frac{1}{a}+\ii\;k}
  \left[1+\frac{\frac{r_0}{2}\;k^2}{\frac{1}{a}+\ii\;k}+\dots\right]
  \Longrightarrow
  \calA(k)=-\frac{4\pi}{M}\;
  \frac{1}{\frac{1}{a}-\frac{r_0}{2}\;k^2+\dots+\ii\;k}
\end{equation} 
Its LO is found for effective range $r_0=0$.  The
strictly-perturbative result on the left provides a small correction as long
as $a\sim\frac{1}{k}\gg r_0\sim\frac{1}{\mpi}$. Its LO pole is at
$k_0=\frac{\ii}{a}$, and shifted\footnote{Yes, poles can be shifted in
  perturbation theory; see for example~\cite{Kaplan:1998we} and also
  formalistic details and references in~\cite{hgrienotes}.} at NLO to
$k_0\approx\frac{\ii}{a}(1+\frac{r_0}{2a})$.

These same poles are also found in the resummed version on the right hand side
at LO ($r_0=0$) and NLO. But there is another NLO pole at
$k_1\approx\frac{2\ii}{r_0}(1-\frac{r_0}{2a})\approx\frac{2\ii}{r_0}\gtrsim\mpi$ with equal but
opposite, and hence unphysical, residue. Since $\frac{r_0}{a}\ll1$, it is
never far from the breakdown scale.  In the \wave{1}{S}{0} channel, for
example, $k_1\approx150\ii\;\MeV\approx\mpi$. At \NXLO{3}, this pole moves to
$130\ii\;\MeV$ and one encounters two more poles at
$[-60\ii\pm350]\;\MeV$. Two more unphysical poles appear at every odd
order. Similarly, the attractive $\frac{1}{r^3}$ part of the tensor one-pion
exchange leads to additional deeply bound states.

Such spurious states not only lead to numerical issues which need nontrivial
solutions, like projecting them out. Partial resummation also softens the
(unphysical) ultraviolet behaviour of the amplitude: The resummed NLO version
of eq.~\eqref{eq:ERE} $\calA_\mathrm{NLO}(k\to\infty)\sim k^{-2}$ converges
more quickly than the LO form, $\calA_\mathrm{LO}(k\to\infty)\sim k^{-1}$,
while the strictly perturbative NLO version is $\sim k^0$. Therefore, if these
amplitudes are inserted into 3N processes, fewer LECs appear to be necessary
to cure residual cutoff dependence at higher orders. In a striking example,
Gabbiani demonstrated that a careless resummation of effective-range
contributions appears to eliminate the need for the very $3\N$ interaction
which is so central to the Efimov effect~\cite{Gabbiani:2001yh}, and that it
also happens to lead to phase shifts which are not supported by data. The
problem might be mitigated by defining a much smaller applicable cutoff window
$\LambdaEFT\lesssim\Lambda\ll k_\mathrm{spur}$. But that makes it much harder
to analyse consistency and cutoff-independence of the EFT power counting. All
power counting developers therefore use strict perturbation theory around a
non-perturbative LO result.

In it, both observables and interactions are expanded in powers of $Q$, and
only matching powers are kept. For example, the NLO correction $T_0$ (blue
hatched) to
eq.~\eqref{eq:consistency} is determined by terms which involve the NLO
interaction $V_0$ once and once only, and the LO $T_{-1}$ (red shaded) only in
half-off-shell kinematics:
\begin{equation}
  \label{eq:nlo}
  \begin{split}
    \includegraphics[clip=,width=0.9\linewidth]{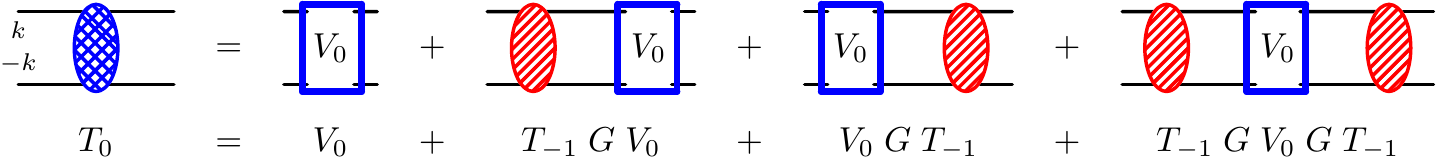}
  \end{split}
\end{equation}
The total amplitude is then the sum of LO and NLO, $T=T_{-1}+T_0$.  Starting
at \NXLO{2}, terms like
\begin{equation}
  \label{eq:nnlo}
  \begin{split}
    T_{-1} \;V_0\;T_{-1}\;V_0\;T_{-1}\;\;:\;\;
    \mbox{
      \parbox{0.28\linewidth}
      {\includegraphics[clip=,width=\linewidth]{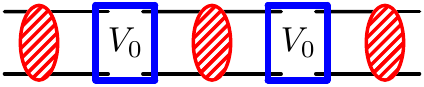}}}
  \end{split}
\end{equation}
appear to require LO amplitudes with both incident and outgoing momenta
off-shell, sandwiched between NLO corrections $V_0$. Many, like me, were happy
with strict perturbation at NLO but thought \NXLO{2} and beyond was just too
much work, turned to partial resummation, and discouraged others to push
further~\cite{Bedaque:1999vb}.

Fortunately, Jared Vanasse~\cite{Vanasse:2013sda} did not listen and in 2013
re-discovered what, embarrassingly, has for over a century been known in
mathematical Perturbation Theory. I even taught it regularly in my
Mathematical Methods class without making the connection. One never needs a
full-off-shell amplitude! The lower-order correction to the half-off-shell
amplitude $T_{i<n}$ and to the interactions $V_{i<n}$ (both red) fully
determine the $n$th correction in an integral equation for the half-off-shell
$T_n$ from term $V_n$ (both blue):
\begin{equation}
  \label{eq:nxlo}
  \begin{split}
  T_n=V_n+\sum\limits_{m=-1}^{n-1}V_{n-m-1}GT_m+V_{-1}GT_n\;\;:\;\;
    \mbox{
      \parbox{0.5\linewidth}
      {\includegraphics[clip=,width=\linewidth]{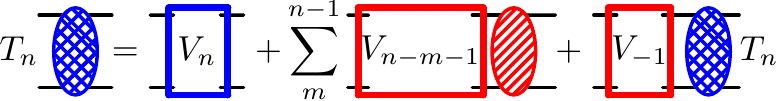}}}
  \end{split}
\end{equation}
This is simply the Distorted-Wave Born Approximation for the
Lippmann-Schwinger equation and implemented quite easily into existing
codes; notes on a number of other approximation methods exist~\cite{hgrienotes}.

\subsection{Listen To Your Elders, Or: Bridging between \EFTNoPion and \ChiEFT}
\label{sec:elders}

The KSW variant of perturbative pions with a non-perturbative contact
interaction at LO~\cite{Kaplan:1998tg,Kaplan:1998we} is a consistent \ChiEFT,
with analytic results in two-nucleon processes. Unfortunately, this beautiful
theory was slain by the ugly fact that the momentum-dependence of the
corresponding counter term at \NXLO{2} in the \wave{3}{SD}{1} wave is not
versatile enough to limit corrections to remain smaller than the leading
pieces beyond momenta $\lesssim200\;\MeV$ or so. It quickly had no resemblance
with data~\cite{Fleming:1999ee}.

Indeed since the inception, pre-EFT people had warned us that in their
experience, the attractive tensor part of one-pion exchange was so strong that
it needed to be iterated to get anywhere near observed phase shifts even at
relatively low energies. I flat-out denied the relevance of that, but
\emph{You don't know what you are talking about}~\cite{Trento}. So we buried
perturbative pions for not living up to their promise.

And yet, there is life in the old dog. Not only are perturbative pions the
consistent \ChiEFT both of the \wave{1}{S}{0} wave~\cite{Beane:2001bc} and of
the \wave{3}{SD}{1} wave at least up to $\mpi$~\cite{Fleming:1999ee}. It is my
firm prejudice, untainted by evidence, that they are also our best hope to
capture the transition from pionless EFT to non-perturbative
pions
.

\subsection{The Thin Blue Line, Or: Reporting Results Needs Theory Uncertainties}
\label{sec:line}

EFTs make a specific promise to facilitate a core tenet of the Scientific
Method: To provide pre- and post-dictions that can be falsified by data. For
that, not only experiments must provide error bars; theorists, too, must
clearly and reproducibly assess their uncertainties, preferably \emph{before}
a closer look at the data to be explained~\cite{editorial}. That means
theorists must provide a probability distribution function encapsulating the
likelihood of their answer, so that its overlap with data can be
quantified. It is insufficient to compare numbers; one also must judge their
reliability. Only then can one enter into an informed scientific discussion
weighing one interpretation against another. A priceless advantage of EFTs is
that its assumptions can be tested ex-post: Are higher orders indeed small? Is
the expansion parameter what it is supposed to be?

Reasonable people can reasonably disagree about to which degree reasonable
assumptions are actually reasonable, but no reasonable dialogue is possible
without disclosing those assumptions in full. Error bars have error
bars~\cite{xkcd}; that is why in modern statistics language, they are called
``confidence intervals''\footnote{\emph{The aim is to estimate the
    uncertainty, not to state the exact amount of the error or provide a
    rigorous bound.}~\cite{editorial}}.

In retrospect, it is astounding how we EFT advocates in the heat on the top
floor of the ECT*'s Rustico in 1999 could endlessly speak of
model-independence, consistency and convergence, while at the same time
showing precious few quantified estimates of EFT truncation
errors~\cite{Trento}. In what surely is a sign of progress, referees now
routinely request a discussion of theory uncertainties. Simply stating that
this is \emph{difficult}~\cite{Trento} is no sufficient excuse any
more~\cite{editorial}. While such discussions were few and far between before,
this past decade saw a barrage of articles on sophisticated tools to
quantitatively test the EFT assumptions. Many of these techniques are
accompanied by software which makes them easy to employ by the average user;
see \eg~\cite{Buqeye, Band}. Bayesian statistical analysis, starting from
reasonable expectations clearly formulated as priors, sets the standard see
\emph{e.g.}~\cite{JPhysG, JPhysG2, Furnstahl:2021rfk, Griesshammer:2020fwr,
  Melendez:2020ikd} and references therein. One can use it to quantify to
which extent the fundamental EFT assumptions actually bear out: order-by-order
convergence; the putative effect of higher-order terms; the values of the
momentum-dependent expansion parameter $Q$ and of the breakdown scale
$\LambdaEFT$; and whether fitted LECs are indeed of natural size, as
Naturalness requires. To be taken seriously, authors have to demonstrate, at
the very least, that what is classified as higher-order terms does indeed
decrease in importance from one order to the next. Bygone the days of plots
with lines of infinitesimal width -- corridors of uncertainties are the New
Normal.  \emph{Real theorists have error bars.}~\cite{Trento}

While a Bayesian interpretation is only as reasonable as its assumptions,
those can, in turn, fortunately be tested for consistency within the formalism
itself, but also outside it. Responsible Scientists are schizophrenic at
heart: both convinced of a result and constantly questioning it at the same
time. The more of our own questions it survives, the more confident we become
-- and sometimes arrogant. So, it is advantageous to query the prior's
posterior by also assessing how stable results are under reasonable variations
not fully captured by Bayesian analysis.

Each of the following methods uses the ``democratic principle'' that
different, reasonable choices must agree up to higher-order corrections. Like
in a democracy, fringe choices will lead to extremist results which should
however be discarded in a healthy discourse. In particular, one can check if
the impact of different choices on observables decrease order-by-order. This
way, one yet again maps out a corridor of theory uncertainties which usually
complement the corridors of Bayesian analyses because they test assumptions
which are at least in part different.

Top of the list is using different numerical cutoffs, and different ways to
regulate, like hard, Gau\3ian and Pauli-Villars cutoff functions.
A dimension-ful cutoff $\Lambda$ has no physical significance; it is merely a
tool to cut off integrals at high momenta or small distances to test to which
degree answers depend on those high-end loop momenta $q\gtrsim\LambdaEFT$
at which the EFT does not capture the correct Physics. That means
$\Lambda\gtrsim\LambdaEFT$ or a bit smaller. While one often talks of
``divergent'' and ``non-renormalised'' answers, this is just short-speech for
``answers which, at any given order, depend more than they should on what
happens at scales at which the theory does not make sense''. But since this is
a mouthful, we use trigger words.

It is tempting to call the result for $\Lambda\to\infty$ ``the'' answer, but
any cutoff is equally legitimate and valid, and none is preferred, as long as
$\Lambda\gtrsim\LambdaEFT$\footnote{Some only vary the cutoff in a window
  around the breakdown scale, $\Lambda\approx\LambdaEFT$, for philosophical or
  numerical reasons.}. This ``democratic principle'' is a neat tool to turn
lines into corridors of uncertainty, even when one is not a purist who
explores a wide cutoff range to develop the correct power counting as in
sect.~\ref{sec:PC}. Fortunately, modern \ChiEFT interactions come with
strengths determined over a reasonable range of cutoffs -- but a wider range
would of course be better. Unfortunately, these also routinely appear to
under-estimate higher-order effects (unless one also samples different cutoff
functions). Therefore, a wide corridor indicates that the question whether or
not data and \ChiEFT agree remains (at best) undecided. But a a narrow
corridor does not make a result highly reliable without additional
corroborating evidence, like a Bayesian uncertainty quantification.

One can also vary the renormalisation point, namely at which energy and from
which observable one determines parameters. For example, the Effective Range
Expansion of eq.~\eqref{eq:ERE} can be about $k=0$, giving the correct
scattering length \etc~Likewise, one can expand about the pole position
$k_0=\frac{\ii}{a}(1+\frac{r_0}{2a})$, so that one starts out with the correct
binding energy even at LO. Indeed, here is another ``democratic principle'':
fitting at any point or corridor $k\ll\LambdaEFT$ is equally legitimate if all
data are of the same quality. Of course, one must avoid $k\nearrow\LambdaEFT$
where the expansion parameter becomes unreliably large.  Overall, fits must be
weighted with $Q$: more constraining at small $Q$, less so as $Q\nearrow1$.
The ``Goldilocks corridor'' $k\sim\ptyp$ is much preferred since it captures
the Physics at the scales the EFT is designed for. User-friendly Bayesian
methods are readily available~\cite{JPhysG,JPhysG2,Furnstahl:2021rfk,
  Melendez:2020ikd}.

\ChiEFT is designed for $k\sim\mpi$, so fits for $k\ll\mpi$ are not as
efficient or meaningful. At these lower scales, fine tuning and universal
correlations take over. For example, the deuteron mean-square radius is
intricately correlated to its binding energy. In the Effective Range Expansion
around $k=0$, eq.~\eqref{eq:ERE}, it moves from $r_\dd=a$ at LO to
$r_\dd=a(1-\frac{r_0}{2a})=\frac{\ii}{k_0}$ at NLO, which also happens to be the
inverse of where the pole of the amplitude is at each of these orders;
\cf~discussion below eq.~\eqref{eq:ERE}. So, one can avoid the trivial
correlation between binding energy $B$ and system size if one starts from the
right binding energy both at LO and NLO:
$r_\dd=\frac{1}{|k_0|=\sqrt{MB}}\pm4\%$. Consequently, a fit to these two data
is highly correlated.  That is but one example of a kinematic point which is,
in a particular context, more important than others.

Likewise, the pion-production threshold is in the cm system at $\mpi$ in LO
\ChiEFT, with nucleon-recoil corrections at higher orders restoring the
kinematically correct position. To avoid trivial correlations, one better
agrees with data on the threshold position by resumming some kinematic
contributions; see \eg~\cite{Griesshammer:2012we}.

One can also include a sub-set of higher-order corrections which by itself
both obeys all symmetries and is not needed to cure other cutoff-dependencies
at the given order (\ie~is renormalisation group invariant by itself).  That
does not include the intrinsic accuracy of the EFT result without additional
\emph{a-priori} justification why to include that particular term, and not
others (see threshold and pole positions above). One just adds some terms one
did not have to add. In return, one can estimate the corridor mapped out by a
democracy of higher-order corrections. In the end, however, the accuracy is
only improved to that of the next order if one includes \emph{all}
contributions at that next order, not just a few which are easy to compute.

All this makes for more work: more computational resources, more thought, more
discussions. But it is time well spent and prevents a quick adrenaline rush of
prematurely claiming victory.

It goes without saying that the corridors of theoretical uncertainties should
honestly be assessed before comparing to data. The more diverse the methods,
the more confident one becomes. After all, we want to determine if the EFT,
with its symmetries, degrees of freedom and power counting, is in
contradiction to Nature or not, namely if data and theory corridors do
consistently and significantly overlap or not. We cannot prove that we are
right, but we can be proven wrong, or become -- with accumulating evidence --
increasingly confident about being right\footnote{\emph{ That is the curse of
    statistics, that it can never prove things, only disprove them!  At best,
    you can substantiate a hypothesis by ruling out, statistically, a whole
    long list of competing hypotheses, every one that has ever been
    proposed. After a while your adversaries and competitors will give up
    trying to think of alternative hypotheses, or else they will grow old and
    die, and then your hypothesis will become accepted. Sounds crazy, we know,
    but that’s how science works!}~\cite[sect.~14.0]{NumRec};
  \cf~sect.~\ref{sec:conclusions}.}. As frustrating as inconclusive results
may be, they are just as worth reporting as agreements or disagreements, so
that others who are smarter can build on them.

\subsection{Variatio Delectat, Or: What EFT Should I Choose?}
\label{sec:variations}

\EFTNoPion, KSW and \ChiEFT both without and with a dynamical Delta are all
perfectly consistent theories of Nuclear Physics. They share symmetries but
contain quite different particles. Each converges order-by-order at some
scale, but the scales where they start to disagree with data is quite
different: $\ptyp\ll\mpi$, $\approx200\;\MeV$, $\Delta_M$, or $700\;\MeV$,
respectively. There are also different proposals to order interactions;
\cf~sect.~\ref{sec:PC}. Likewise, ``non-relativistic'' versions add
relativistic corrections as perturbations and exist alongside
kinematically\footnote{As none contain nucleon--anti-nucleon loops, they are
  not covariant Quantum Field Theories -- merely covariant Quantum Mechanics,
  which is well known to be conceptually inconsistent~\cite{Klein}.} covariant
ones.  Finally, people determine parameters from different data, fit regions
and cutoffs.

This proliferation of variants is confusing -- is \ChiEFT now simply a
collection of different models (sometimes merely of different geographical
origin) which vie for supremacy, often with less-than convincing logical
arguments?  Are we back to \emph{my model is better than yours}~\cite{Trento}?

Yes. No. \ChiEFT is not a fixed set of numbers and interactions but a set of
operational instructions. Any two variants which are known to be consistent
and applicable in some kinematic overlap must agree to within their respective
levels of accuracy inside that region. However, some are technically
simpler\footnote{In QCD, the $\overline{\mathrm{MS}}$ scheme is popular
  because it is simple, but its convergence pattern is actually not that
  stellar.}, converge more quickly (order-by-order, not to data), avoid fine
tuning, or even need fewer parameters at the same level of accuracy. Where a
bouquet of variants overlaps, it estimates yet again residual uncertainties.

The variant to choose depends thus also on the problem's scale and required
accuracy -- and, most importantly, on what question to answer. Interactions
fitted to NN and few-N data robustly explain the gross structures of heavy
nuclei. But to get more details right, one better uses parameters from, say,
light nuclei. EFTs build bridges between simpler and more complex systems, and
explain patterns in the latter.

However, it is not ``anything goes''. As much as one might be tempted, one
can\footnote{One ``can'' (is able to), but one ``should'' not (is not allowed
  to).}, for example, not sandwich pionless interactions whose strengths are
determined in \EFTNoPion between wave functions of \ChiEFT to compute nuclear
matrix elements. These two theories have an entirely different particle
content (pions or not), in part even different symmetries, and definitely
different short-distance behaviour, so the result is inherently unstable
against variations of the cutoff. Such mix-and-match can only be compared to
wearing a pair of red trousers with a polka-dot shirt.


\subsection{Endgame, Or: What Do We Want To Achieve?\protect\footnote{For the
    thoughts in this subsection, I am particularly indebted to M.~Savage's
    moderation of a discussion
    at the 2015 ECT* workshop \textsc{New Ideas in
      Constraining Nuclear Forces} with uncomfortable and therefore
    thought-provoking questions.}}
\label{sec:details}

What is the goal of an EFT of Nuclear Physics?  EFTs are bridges between the
microscopic and the macroscopic -- \ChiEFT is one from the quark-gluon version
of QCD both to \EFTNoPion and to Nuclear Structure. They aim to explain
relations and structures and are set to perform reasonably in the overwhelming
majority of tests. \ChiEFT will not precisely predict the intricate energy
spacing and ordering in Linoleum-314~\cite{Trento}; if necessary, we have
another EFT for that.  Indeed, EFTs do not attempt to describe all aspects of
the real world at a given scale completely. They cannot be beaten by ``nuclear
engineering'', namely models fine-tuned to particular details of particular
systems, at the cost of failing in almost all other
situations\footnote{\emph{In fact the trouble in the recent past has been a
    surfeit of different models \emph{[of the nucleus]}, each of them successful in
    explaining the behavior of nuclei in some situations, and each in apparent
    contradiction with other successful models or with our ideas about nuclear
    forces.}~\cite{Peierls}}.

A central EFT promise is that it encodes the unresolved short-distance
information at a given accuracy into not just some, but the
\emph{smallest-possible} number of independent LECs constrained by a set of
symmetries. If a set we thought should be minimal shows actually
correlations, then the theory is not yet reduced to its minimal information
content, for example because we are missing pivotal symmetries;
\cf~sect.~\ref{sec:symmetries}.

That maximally-compressed information is what survives as important to the
low-resolution version of the high-resolution theory. And that is why counting
powers and finding the smallest-possible number of parameters is so
imperative. If two EFTs are renormalised and describe the same data with the
same accuracy, the one with the least number of parameters wins because it
needs the least information.

But trust in EFT methods and methodologies must be earned by demonstrating
that results agree with Nature at least for a few ``signal observables'':
non-trivial data which have ideally both eluded explanations in the past and
are of great importance, for example for key astrophysical processes which
help us interpret our place in the Universe. The general public that so
gracefully and patiently finances our passion can expect the community to
formulate overarching goals and to coordinate the effort towards them. While
one may be reluctant to declare success, whatever that means, in
``explaining'' Nuclear Physics, it is important to do so for at least some
subsections we care about.

\subsection{Gripes of Wrath}
\label{sec:gripes}

One could write about many more mistakes and things we are doing better now
that before Weinberg's contributions. For decades, we have used few-N
potentials which largely neglect retardation effects in pion exchange (some
recent work includes it perturbatively). For decades, we fit to $\N\N$ data
above the pion-production threshold with potentials that do not allow for pion
production (some recent work stops the fitting just below the pion-production
threshold).

Despite two decades of complaints, we also still do not have a common standard
to share code or results, be it potentials, interactions, wave functions or
matrix elements, so that our peers can use them as input. There are a few but
notable exceptions, including the \texttt{github} repositories of the
\textsc{Buqeye}~\cite{Buqeye} and \textsc{Band}~\cite{Band} collaborations,
the self-consistent Green's function code by Barbieri~\cite{Barbieri},
Stroberg's in-medium Similarity Renormalisation Group code~\cite{Stroberg} and
shell model codes like \textsc{NuShellX}~\cite{nushellx},
\textsc{KShell}~\cite{kshell}, \textsc{Bigstick}~\cite{bigstick} and
\textsc{AntOine}~\cite{antoine}. That lattice-QCD is mandated to make
configurations and codes fully available after an embargo period, has made the
groundbreaking research by upstarts like NPLQCD possible. We know that coding
needs dedicated experts, but we have no agreed mechanism to credit them and
reward their dedication such that they feel safe to invest the work to
document their codes and make them public, without fear of the question what
``Actual Physics'' they accomplished. On both these issues, see also a recent
memorandum on an open-source toolchain for \emph{ab initio} Nuclear
Physics~\cite{perspectives}.

We continue what we have done so far because it is less dangerous and produces
more publications per year than changing course\footnote{One of the referees
  reminds me, however, that we sometimes continue what we have done so far
  because we actually believe in it.}. We set publications aside which
introduce conceptual ideas but do not immediately find a killer
application. We are risk-averse because articles about what did not work are
difficult to get published. We spin even defeats into victories.
%
%
We are doing a million things wrong, and only a few middling-right, amongst
them that we listen and learn -- sometimes.

\section{Oh Now That's a Blueprint for an Impossibly Rosy Future\protect~\cite{Maguire}}
\label{sec:conclusions}

Max Planck had a very dark take on scientific progress, often condensed into
the dictum \emph{Science advances one funeral at a time}\footnote{\emph{Eine
    neue wissenschaftliche Wahrheit pflegt sich nicht in der Weise
    durchzusetzen, da\3 ihre Gegner \"uberzeugt werden und sich als belehrt
    erkl\"aren, sondern vielmehr dadurch, da\3 ihre Gegner allm\"ahlich
    aussterben und da\3 die heranwachsende Generation von vornherein mit der
    Wahrheit vertraut gemacht ist.}~\cite{Planck}}. But despite all our
shortcomings, Nuclear Physics has revolutionised and re-invented itself in the
thirty years since Weinberg's foundational
contributions~\cite{Weinberg:1990rz, Weinberg:1991um, Weinberg:1992yk}. Jim
Friar's prediction in 1999 became true, he was just too optimistic by a factor
of ten: \emph{In 1-to-2 years, we will all be using $\chi$PT-designed
  products.}~\cite{Trento} In this issue, Ruprecht Machleidt eloquently
describes how, after initial scepticism, the Nuclear community not just
adapted and adopted, but embraced \ChiEFT~\cite{Machleidt:2021ggx}.

On the way, we doled out foolish advice and ridiculed sage one. In the next
thirty years and beyond, we will forget lessons learned, indulge in new
mistakes, explore new cul-de-sacs and commit new follies. To crawl or even
more-than-crawl forward on the road of progress will continue to need
\emph{blood, toil, tears and sweat}~\cite{Churchill}. We do Nuclear Physics
not because it is easy, but because it is hard -- but hopefully, new
generations will make light of present and past struggles as
trivial~\cite{Crusher}.

\begin{flushright}
  \emph{Wrong theories are not an impediment to the progress of science.\\
    They are a central part of the struggle.}~\cite{Dyson}
\end{flushright}

\section*{Data Availability Statement}

The preciously few data underlying this work are available in full upon
request from the author, but identifying information about individual quotes
will not be surrendered.

\begin{acknowledgements}
  It is a pleasure to thank all combatants of the 1999 \textsc{Workshop on
    Nuclear Forces} at the European Centre for Theoretical Studies in Nuclear
  Physics and Related Areas ECT* in Trento (Italy) who proved that one can
  have both more-than-frank exchanges of ideas during the day and share good
  food and \emph{gelato} around the Piazza at night. 
  Even an approximate list of the most important individuals whose thoughts
  and conversations with me and others have shaped the views expressed here is
  too long for these margins: Thank you, and you are not responsible for my
  mistakes.
  J.~Kirscher suffered through a draft without major harm. A.~Long's detailed
  notes were an appeal for more examples and more clarity on behalf of those
  who had not been in the line of this kind of work for two or more decades,
  and were seconded by additional communications from the community.
  D.~R.~Phillips' and J.~de Vries' unsolicited feedback closed gaping holes.
  This work was supported in part by the US Department of Energy under
  contract DE-SC0015393 and conducted in part in GW's Campus in the Closet.
\end{acknowledgements}


\end{document}